\documentclass[journal ]{new-aiaa}
\usepackage[utf8]{inputenc}
\usepackage{textcomp}

\usepackage{graphicx}
\usepackage{amsmath}
\usepackage[version=4]{mhchem}
\usepackage{longtable,tabularx}
\usepackage[per-mode=symbol]{siunitx}
\usepackage{hyperref}
\usepackage[ruled,vlined,linesnumbered]{algorithm2e}
\usepackage{float}
\usepackage{array} 
\usepackage{subcaption}
\usepackage{siunitx}

\title{An Euler–Lagrangian Multiphysics Coupling Framework \\ for Particle-Laden High-Speed Flows}

\author{Hyeon Woo Nam\footnote{Graduate Student, Department of Aerospace Engineering, hyeon.aero@kaist.ac.kr.}}
\author{Tae Woong Jeong\footnote{Graduate Student, Department of Aerospace Engineering.}}
\affil{Korea Advanced Institute of Science and Technology, Daejeon 34141, Republic of Korea}

\author{Sung Min Jo\footnote{Assistant Professor, Department of Mechanical and Aerospace Engineering, sungmin.jo@ucf.edu (Corresponding Author).}}

\affil{UCF Center of Excellence in Hypersonics and Space Propulsion (HYPERSPACE), \\ University of Central Florida, Orlando, FL 32816, USA}

\begin{document}

\maketitle

\begin{abstract}
Particle-laden effects in high-speed flows require a coupled Euler and Lagrangian prediction technique with varying fidelity of thermochemical models, depending on the simulation conditions of interest. This requirement makes the development of a conventional monolithic solver challenging to manage the different fidelity of the thermochemical models within a single computational framework. To address this, the present study proposes a multi-solver framework for the coupled Euler-Lagrangian predictions applicable to various particle-laden high-speed flow conditions. Volumetric and surface couplings are established between a particle solver ORACLE (OpenFOAM-based lagRAngian CoupLEr) and a thermochemical nonequilibrium flow solver based on an adaptable data exchange algorithm. The developed framework is then validated by predicting particle-laden supersonic nozzle flows and aerothermal heating around a hypersonic Martian atmospheric entry capsule. Finally, a quasi-1D approximation is proposed in conjunction with a surrogate method to efficiently and accurately predict particle-laden surface erosion, with quantified parametric uncertainty, for hypersonic aerothermal characterization.

\end{abstract}

\section{Introduction}
\label{Introduction}
High-speed particle-laden flows are an essential concern in various aerospace applications. In supersonic nozzles, particles from incomplete combustion create two-phase flows that alter the internal flow field and change the shock strength and location \cite{sahai2022variable}. During retropropulsion, dust interacts with supersonic plumes, potentially damaging the lander’s sensors and onboard hardware \cite{berger2013role}. For the hypersonic Entry Descent and Landing (EDL) phase to Mars, the impacts of atmospheric dust particles and thermochemical reactions on the vehicle surface lead to the recession of the thermal protection system (TPS) \cite{palmer2020modeling,palmer2000reassessment, jo2020stagnation}. In addition to the mechanical damage, the atmospheric dust particles undergo force and heat transfer interactions while traversing the shock layer, resulting in boundary layer thickness modifications that cause vehicle heat flux to deviate from pure gas predictions \cite{padmapriya2001numerical, wang1988compressible, hinkle2022efficient}.

Addressing the multiphysics phenomena related to particles requires an Euler-Lagrangian framework capable of resolving particle forces and heat transfer at high-speeds under both thermochemical equilibrium and nonequilibrium conditions. Previous numerical studies have explored various strategies for modeling high-speed dusty flows. \citet{palmer2020modeling} employed a Monte Carlo method to investigate thermochemical reactions and dust erosion along Martian entry trajectories based on a particle size distribution model. \citet{sahai2022variable} developed a computational fluid dynamics–discrete element method (CFD–DEM) framework with adjustable fidelity, ranging from one-way to four-way coupling for high-speed flows. \citet{ching2020two} applied a high-order Galerkin scheme for Lagrangian particle tracking. In contrast, \citet{hinkle2024efficient} introduced the Trajectory Control Volume (TCV) method, which defines reduced control volumes along particle paths to compute source terms and particle–surface interactions, assuming flux conservation. The TCV approach significantly reduces computational costs compared with Monte Carlo methods when particle dynamics remain relatively simple. Most previous studies adopt a monolithic approach in which solvers for each physical domain are tightly integrated in a dependent manner.

When developing a multiphysics coupled framework, conventional monolithic approach may present limitations in modularity, verification, and maintenance; therefore, the present work employs a multi-solver coupling approach that integrates independent solvers. This modular approach facilitates development, validation, and maintenance by isolating individual physical domains. The Eulerian domain utilizes HEGEL (High-fidElity tool for maGnEto-gasdynamics simuLations) \cite{munafo2024hegel}, a high-fidelity thermochemical nonequilibrium flow solver, along with a thermophysical library, PLATO (PLAsmas in Thermodynamic nonequilibrium) \cite{munafo2025plato}. The Lagrangian domain employs ORACLE (OpenFOAM-based lagRAngian CoupLEr), a newly developed particle solver based on OpenFOAM \cite{weller1998tensorial} that models discrete particle dynamics in high-speed flow regimes. The Eulerian and Lagrangian fields are then linked by defining a coupling domain for volumetric and surface data exchange.

The objectives of this work are twofold. The first objective is the development and validation of ORACLE, a high-speed particle flow solver, and the Euler-Lagrangian coupling framework. Validation cases include the Jet Propulsion Laboratory (JPL) supersonic converging-diverging nozzle \cite{back1966detection} and the ExoMars Schiaparelli capsule entry \cite{palmer2020modeling}, representing typical two-way coupling validation cases. The second objective is to investigate surface recession due to particle impact during the Schiaparelli capsule entry into the Martian atmosphere along flight trajectories using the developed framework. While previous studies conducted 3D simulations with an angle of attack, this work proposes a quasi-1D approach as a cost-effective alternative that facilitates extensive investigations of parameteric uncertainty.

In the early design stage, a prediction of the recession rate is needed to determine the required TPS thickness \cite{boland2023dust, le2025numerical}. Previous efforts attempted to develop engineering correlations for recession rate; however, these correlations were limited to spherical bodies and showed prediction errors within about 16\% in all cases \cite{boland2023dust}. In this study, instead of relying on the simplified correlations, a data-driven surrogate model for the recession rate is proposed based on detailed simulations of realistic geometries to enable fast and accurate predictions of the particle-induced recession rate.

The remainder of this paper is organized as follows. Section~\ref{Euler-Lagrangian Coupling Framework_Section} describes the developed Euler-Lagrangian coupling framework, including the governing equations and physical modeling for each domain. Section~\ref{Results and Discussion} details validation against the JPL supersonic converging–diverging nozzle and the ExoMars Schiaparelli entry. Section~\ref{Reduced-Order Modeling of Cumulative Surface Recession} presents a quasi-1D analysis using the validated framework and demonstrates a proposed data-driven surrogate model of particle-induced surface recession along the flight trajectory. Section~\ref{Conclusion} provides conclusions. The \hyperref[Appendix]{Appendix} formulates basis functions and coefficients for the proposed surrogate model that the relevant research community can employ.

\section{Euler-Lagrangian Coupling Framework}
\label{Euler-Lagrangian Coupling Framework_Section}

\subsection{Coupling Strategy}

\begin{figure}[H]
\centerline{\includegraphics[width=8.0cm]{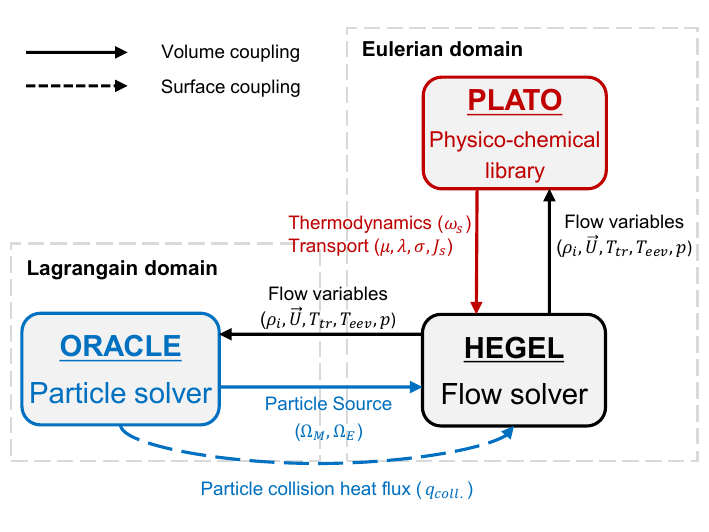}}
    \caption[Proposed Euler–Lagrangian coupling framework for particle-laden high-speed flows.]{Proposed Euler–Lagrangian coupling framework for particle-laden high-speed flows.} \label{Euler-Lagrangian coupling framework}
\end{figure}

In this study, a coupling framework was developed to analyze multiphysics-coupled multiphase hypersonic flows. Figure~\ref{Euler-Lagrangian coupling framework} presents an overview of the coupling framework, consisting of the solvers, PLATO, HEGEL, and ORACLE. Data exchange among the solvers relies on a coupling library, preCICE \cite{bungartz2016precice}. Through the coupling, each solver exchanges the relevant properties with the corresponding domains of the other solvers at a predefined data exchange frequency.

HEGEL supplies flow variables such as gas phase velocity, pressure, and density to ORACLE. ORACLE uses these inputs to compute particle dynamics via the force/heat transfer models and governing equations. As particles traverse computational cells, they contribute momentum and energy back to the continuous phase; thus, the solution of the particle equations provides source terms of the gas phase conservation equations. Additionally, when particles collide with a surface, ORACLE calculates the resulting surface collision heat flux and returns it to HEGEL to update the surface boundary conditions. By ensuring consistency in the quantities exchanged, conservation is maintained across the interface.

\newpage
Algorithm~\ref{algorithm1} further details the HEGEL--ORACLE two-way coupling procedure, including the exchange of gas properties such as 
$T_{i,g} \in \{ T_{\mathrm{tr}}, \, T_{\mathrm{eev}} \}$ under thermochemical equilibrium and nonequilibrium conditions. $T_{\mathrm{tr}}$ and $T_{\mathrm{eev}}$ represent translational-rotational temperature and electron-electronic-vibrational temperature, respectively. The subscripts $g$ and $p$ refer to the gas phase and particle phase. The gas phase field properties $( \boldsymbol{v}_g, p_g, \rho_g)$ denote the velocity, pressure, and density of the gas. The terms $\Omega_{M,p}$ and $\Omega_{E,p}$ are momentum and energy source terms resulting from the particle-fluid interactions, computed based on the particle properties $(m_p,d_p,\boldsymbol{x}_p,\boldsymbol{v}_p, T_p, \rho_p)$, which represent the particle mass, diameter, position, velocity, temperature, and density, respectively. The Eulerian (HEGEL) and Lagrangian (ORACLE) solvers advance on independent time scales and exchange data at a user-defined coupling intervals. At each interval, the flow solver iteratively converges to an updated field, and the particle solver advances to capture the particle dynamics. This process repeats until the HEGEL residual tolerance is satisfied or the final time is reached. The coupling strategy follows a parallel–explicit scheme, in which both solvers progress concurrently and exchange data explicitly at each coupling point with a single sub-iteration\cite{bungartz2016precice}.

\begin{algorithm}[H]
\SetAlgoNoLine
\label{algorithm1}
\caption{HEGEL-ORACLE Two-Way Coupling}
\SetKwInput{KwInput}{Input}
\SetKwInput{KwOutput}{Output}
\KwInput{coupling interval $\Delta t_{c}$, HEGEL time step $\Delta t_g$, ORACLE time step $\Delta t_p$ and HEGEL tolerance $\varepsilon_g$}
\KwOutput{%
  updated flow field solution and source terms for particles $\{\boldsymbol{\Omega}_{M,p},\Omega_{E,p}\}$%
  
}
\DontPrintSemicolon

\textbf{Initialize:} obtain initial flow-field solution and set particle source terms $\{\boldsymbol{\Omega}_{M,p},\Omega_{E,p}\} = \{\boldsymbol{0},0\}$;

\While{$t < t_{\mathrm{final}}$}{
  update coupling time, $t_{c} \leftarrow t_c+\Delta t_{c};$\;

  \While{$t_g < t_c$}{
    solve the (nonequilibrium/equilibrium) Navier–Stokes equations with $\{\boldsymbol{\Omega}_{M,p},\Omega_{E,p}\}$;\; 
    update solver time, $t_g \leftarrow \min\{t_g + \Delta t_g,\, t_c\};$\;
  }
  \While{$t_p < t_c$}{
  \ForEach{\textnormal{particle}}{
    \While{$\boldsymbol{x}_p \in \mathcal{D}$ \textnormal{(Computational domain)} }
    {
        interpolate gas properties at particle position, $\boldsymbol{x}_p$;\;
        solve governing equations for particles over $\Delta t_p$;\;
        update particle properties, $(\boldsymbol{x}_p,\boldsymbol{v}_p,T_{p},d_p,m_p)$;\;
        accumulate $\{\boldsymbol{\Omega}_{M,p},\Omega_{E,p}\}$ in the owner cell;\;
    }
}
    
    update solver time, $t_p \leftarrow \min\{t_p + \Delta t_p,\, t_c\};$\;
  }
\textbf{HEGEL:} write flow variables $\{\boldsymbol{v}_g,p_g,\rho_g,T_{i,g}\}$ and read particle sources $\{\boldsymbol{\Omega}_{M,p},\Omega_{E,p}\};$\;
\textbf{ORACLE:} write particle sources $\{\boldsymbol{\Omega}_{M,p},\Omega_{E,p}\}$ and read flow variables $\{\boldsymbol{v}_g,p_g,\rho_g,T_{i,g}\};$\;
\If{\textnormal{HEGEL residual} $\le \varepsilon_g$}{\textbf{break};}
update global time, $t \leftarrow t_c$;\;
}
\textbf{Finalize:} terminate the coupling.
\end{algorithm}

\subsection{Lagrangian Particle Solver for High-Speed Flows}
\subsubsection{Physical Modeling}

In the present study, particle dynamics in high-speed environments are modeled through the development of ORACLE. Particles are assumed to be spherical point-particles, and their translational motion is tracked along their trajectories. The governing equations are expressed in terms of the commonly used primitive variables, and the particle diameter, momentum, and energy are updated \cite{palmer2020modeling,hinkle2024efficient,ching2020two}. Under thermochemical nonequilibrium conditions, the governing equations can be expressed as follows \cite{hinkle2024thesis}:

\begin{equation}
\label{eq:diameter}
\frac{\partial d_p}{\partial t} = \frac{2\dot{m}_p}{\pi d_p^2 \rho_p},
\end{equation}

\begin{equation}
\label{eq:momentum}
\frac{\partial \boldsymbol{v}_p}{\partial t} 
= \frac{\boldsymbol{F}_p}{m_p}
= \frac{\rho_g C_{D} A_p | \boldsymbol{v}_g - \boldsymbol{v}_p | (\boldsymbol{v}_g - \boldsymbol{v}_p)}{2 m_p},
\end{equation}

\begin{equation}
\label{eq:energy}
\frac{\partial T_p}{\partial t} 
= \frac{\dot{Q}_p}{m_p c_{p,p}} 
= \frac{\pi d_p Nu_g }{m_p c_{p,p}}[\kappa_{tr,g}(T_{tr,g}-T_p ) +\kappa_{eev,g}(T_{eev,g}-T_p ) ] .
\end{equation}

\noindent Here, $C_D$ denotes the drag coefficient, $Nu$ the Nusselt number, $A_p = \pi d_p^2/4$ the particle cross-sectional area, and $c_{p,p}$ the specific heat at constant pressure of the particle. The terms $k_{tr,g}$ and ${k_{eev,g}}$ represent the translational-rotational and electron-electronic-vibrational thermal conductivities of the gas, respectively. The particle force vector $\boldsymbol{F}_p$ is modeled under the quasi-steady assumption, accounting only for aerodynamic drag. The particle heat flux $\dot{Q}_p$ is evaluated using convective heat transfer correlations. During framework validation, the present solver employs drag and heat transfer correlations to ensure consistency with previous studies. For the particle drag coefficient, the correlation of \citet{clift2005bubbles} is applied in continuum regimes, while \citet{henderson1976drag} is used for thermochemical nonequilibrium and rarefied flows. For the particle Nusselt number, the correlation of \citet{drake1961discussion} is employed under equilibrium conditions, and the model of \citet{fox1978shock} is applied in thermochemical nonequilibrium regimes. All these models provide drag and heat transfer coefficients for spherical particles immersed in the surrounding flow.

In addition, particle vaporization is accounted for in high-temperature regions. Following previous studies \cite{palmer2020modeling,hinkle2022efficient}, when the particle surface temperature exceeds the vaporization temperature \cite{schaefer2004thermodynamic}, the absorbed heat flux is assumed to cause mass loss, leading to a decrease in particle radius over time.

\subsubsection{Integration of Particle Effects for Multiphysics Coupling}
\label{source}

In the present work, a two-way coupled approach is employed, while inter-particle collisions are neglected due to the low mass loading of the flow conditions of interest \cite{michaelides2022multiphase,ching2021development}. Based on the tracked particle properties obtained from the governing equations, cumulative source terms are constructed to represent the effects of particles within the domain. These source terms modify the momentum and energy fields of the carrier gas:

\begin{equation}
\boldsymbol{\Omega}_{M,p} = - \frac{1}{V_{\text{cell}}}\sum_{i=1}^{n_p} \boldsymbol{F}_p^i,
\end{equation}

\begin{equation}
\Omega_{E,p} = - \frac{1}{V_{\text{cell}}} \sum_{i=1}^{n_p} \left( \dot{Q_p}^i + \boldsymbol{F}_p^i \cdot \boldsymbol{v}_p^i \right),
\end{equation}

\noindent
where $\boldsymbol{\Omega}_{M,p}$ and $\Omega_{E,p}$ denote the particle-induced source terms for momentum and energy, respectively. $n_p$ is the total number of particles in the owner cell, and $V_{cell}$ is the corresponding cell volume. The coupling between particles and the flow field is achieved through these source terms. Although the vaporization of particles is considered at high temperatures, the resulting products are neglected in the gas phase \cite{hinkle2024efficient}. Therefore, mass exchange between the gas and particles is not considered in this study.

\subsubsection{Particle–Wall Interaction}
Particle–wall interaction is modeled through two mechanisms: (i) surface collision heating and (ii) surface recession. Collision heating is evaluated by assuming that the entire kinetic energy of the incident particle is converted into heat upon impact \cite{ching2022simulations, ching2020two}. The resulting collision heat flux is added to the convective contribution to form the total heat flux, which is applied as a boundary condition to update the surface temperature.

In addition to the collision heating, surface recession is modeled for the Norcoat Liège TPS  of the Schiaparelli capsule. Each impacting particle is assumed to form a hemispherical crater, from which the local recession depth is calculated as follows \cite{palmer2020modeling,keller2009dust}: 

\begin{equation}
\label{eq:penetration_depth}
{P} = 0.00028\rho_p^{0.62} d_p^{1.04867} \lvert \boldsymbol{v}_{p} \rvert^{0.6657},
\end{equation}

\noindent where $P$ denotes the penetration depth, and the crater diameter is defined as $D_c = 2 P$ under the hemispherical assumption. The cumulative surface recession $\Sigma P$ is then obtained by scaling with the number of particles per unit area $\bar{N}_p$ and summing the contributions of all striking particles:

\begin{equation}
\label{eq:local_recession}
\Sigma P \approx \sum D_{c}^{2}{P}\cos\left(\theta\right)\bar{N}_p .
\end{equation}

\begin{figure}[H]
\centerline{\includegraphics[width=8.0cm]{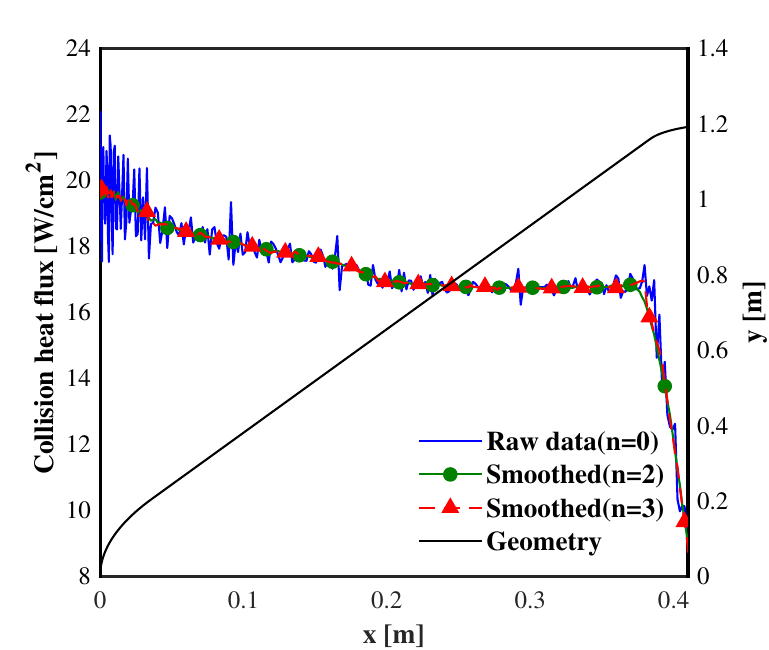}}
    \caption[Effect of the spatial moving-average filter on the surface collision heat flux.]{Effect of the spatial moving-average filter on the surface collision heat flux.} \label{Effect of the spatial moving-average filter on the surface collision heat flux.}
\end{figure}

\noindent
The $\cos(\theta)$ term accounts for the geometric effect of crater shape, where the angle $\theta$ between the particle velocity vector and the surface normal vector is assumed to be 45 degrees \cite{papadopoulos1993heatshield}. When particles are injected across the entire domain using a Monte Carlo approach, statistical noise \cite{kroells2025investigation} arises in both surface heating and recession. To mitigate this effect, a spatial moving-average filter is applied to the computed surface quantities. For a target face $j$, the filtered value $\bar{q}_j$ is defined as

\begin{equation}
\bar{q}_j = \frac{1}{2n+1} \sum_{k=-n}^{n} q_{j+k},
\end{equation}

\noindent where $q_j$ is the unfiltered value, $n$ is the number of neighboring faces on each side of the target face, and $k$ is the index offset of neighboring faces. 
Figure~\ref{Effect of the spatial moving-average filter on the surface collision heat flux.} illustrates the collision heat flux for Exomars Schiaparelli case as discussed in Section~\ref{Schiaparelli}, demonstrating the effect of the moving-average filter. Comparing averaging widths ($n = 0, 2, 3$) shows that $n = 2$ effectively suppresses noise, while a further increase to $n = 3$ provides no additional improvement. Therefore, $n = 2$ is adopted in this study.

\subsection{Thermochemical nonequilibrium Flow Solver}

In this study, an in-house CFD solver HEGEL \cite{munafo2024hegel} is employed. HEGEL is a cell-centered finite volume flow solver capable of solving both thermochemical equilibrium and nonequilibrium models. Using PLATO \cite{munafo2025plato}, HEGEL enables thermophysical analyses in various fidelities, focusing on high-speed flow problems such as hypersonic atmospheric entry cases.  
The governing equations for a thermochemical nonequilibrium two-temperature model are given as follows:

\begin{equation}
\frac{\partial}{\partial t} \left( \rho_{i,g} \boldsymbol{v}_g \right )
+ \nabla \cdot \left( \rho_{i,g} \boldsymbol{v}_g \boldsymbol{v}_g \right)
+ \nabla p_g - \nabla \cdot \boldsymbol{\tau}_g = \boldsymbol{\Omega}_{\text{M},p},
\end{equation}

\begin{equation}
\frac{\partial}{\partial t} \left( \rho_{i,g} E_{g} \right) 
+ \nabla \cdot \left( \rho_g H_g \boldsymbol{v}_g \right) 
+ \nabla \cdot \left( \sum_i \rho_{i,g} h_{i,g} \boldsymbol{V}_{i,g} \right) 
= \nabla \cdot \left( \boldsymbol{\tau}_g \cdot \boldsymbol{v}_g \right) 
- \nabla \cdot \boldsymbol{q}_g + \Omega_{\text{E},p},
\end{equation}

\begin{equation}
\frac{\partial}{\partial t} \left( \rho_g e_{\text{eev},g} \right)
+ \nabla \cdot \left( \rho_g e_{\text{eev},g} \boldsymbol{v}_g \right)
+ \nabla \cdot \left( \sum_i \rho_{i,g} h_{\text{eev},i,g} \boldsymbol{V}_{i,g} \right)
= -p_{e,g} \nabla \cdot \boldsymbol{v}_g
- \nabla \cdot \boldsymbol{q}_{\text{eev},g}
 + \Omega_{\text{ET},g} + \Omega_{\text{CE},g} + \Omega_{\text{VT},g}.
\end{equation}

\noindent
where $\rho_{i,g}$ is the density of species $i$, and $\boldsymbol{\tau}_g$ denotes the viscous stress tensor. $h_{i,g}$ and $\boldsymbol{V}_{i,g}$ are the enthalpy and diffusion velocity vector of species $i$. $\boldsymbol{q}_g$ represents the conductive heat flux vector. 
$e_{\text{eev},g}$ and $h_{\text{eev}, i,g}$ are the mixture and species electron-electronic-vibrational energy and enthalpy, respectively. $p_{e,g}$ and $\boldsymbol{q}_{\text{eev},g}$ denote the electron pressure and the electron-electronic-vibrational heat flux. $H$ and $E$ represent total energy and enthalpy, respectively. The species continuity equation is not shown here because it is not affected by the Euler-Lagrangian coupled analysis of the present study, as discussed in Section \ref{source}. Further details of the thermochemical nonequilibrium flow governing equations can be found in \cite{munafo2020computational, jo2023multi}. A second-order finite volume approach with the Van Leer convective flux scheme \cite{van2005flux} is adopted for the discretization, and a local time-stepping procedure is applied for steady-state computations. The computational domain is constructed using a structured mesh topology. The source terms on the right-hand side are supplied by the other solvers in the framework:  
The electron-translational ($\Omega_{\text{ET},g}$), vibrational-translational ($\Omega_{\text{VT},g}$), and chemistry-electronic ($\Omega_{\text{CE},g}$) energy transfer terms are computed by PLATO. The effects of particles, analyzed by ORACLE, are included in the momentum and total energy equations as particle momentum ($\boldsymbol{\Omega}_{M,p}$) and energy ($\Omega_{E,p}$) source terms, respectively.

\section{Results and Discussion}
\label{Results and Discussion}

\subsection{Particle-Laden Converging-Diverging Nozzle Flow}
A test case based on the converging-diverging nozzle geometry designed by the JPL \cite{back1966detection} is considered. Firstly, results of the pure gas case are compared with both experimental \cite{back1966detection} and previous numerical \cite{sahai2022variable, ching2020two, chang1996application, moukalled2003pressure} studies to establish a baseline, after which the dusty gas flow results are compared with previous numerical studies to verify the two-way coupling framework. The JPL nozzle geometry is described in the work by Back and Cuffel \cite{back1966detection}, and the particle properties and flow conditions are summarized in Table \ref{tab:nozzle-properties}. As in the previous studies, the drag coefficient and heat transfer models used are given by Clift \cite{clift2005bubbles} and Drake \cite{drake1961discussion}, respectively. The test gas is air, and transport properties are modeled using Sutherland's law \cite{sahai2022variable}, as the flow can be approximated as a perfect gas in this case due to the low temperature and moderate flow Mach number. The simulation is performed under axisymmetric assumptions, with a mesh consisting of 186 cells in the axial direction and 59 cells in the radial direction. For the boundary conditions, the inlet is set to a chamber condition, the wall is modeled as a slip wall, and symmetry is applied along the axis. The outlet is treated with a zero-gradient boundary condition, in which flow variables are extrapolated from the interior.

\begin{table}[H]
\centering
\caption{Gas and particle phase properties for the JPL nozzle case \cite{back1966detection}.}
\label{tab:nozzle-properties}
\renewcommand{\arraystretch}{0.8}
\begin{tabular}{cccccccc}
\hline\hline
$P_{c,t}$ (Pa) & $T_{c,t}$ (K) & $Pr$ & $c_{p,g}$ (J/kg$\cdot$K) &
$c_{p,p}$ (J/kg$\cdot$K) & $\rho_p$ (kg/m$^3$) & $d_p$ ($\mu$m) & $\beta$ \\
\hline
$1.034\times10^{6}$ & 555 & 0.72 & 1070 & 1380 & 4004.62 & 20 & 0.3 \\
\hline\hline
\end{tabular}
\end{table}

\begin{figure}[H]
\centering
\begin{subfigure}[H]{0.5\textwidth}
    \centering
    \includegraphics[width=\textwidth]{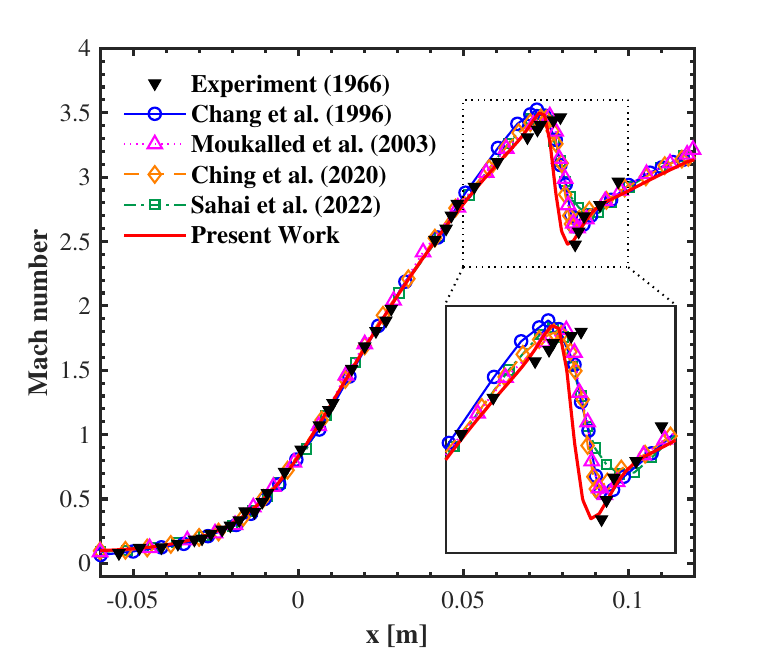}
    \caption{}
    \label{fig:JPL_pure_centerline}
\end{subfigure}%
\hfill
\begin{subfigure}[H]{0.5\textwidth}
    \centering
    \includegraphics[width=\textwidth]{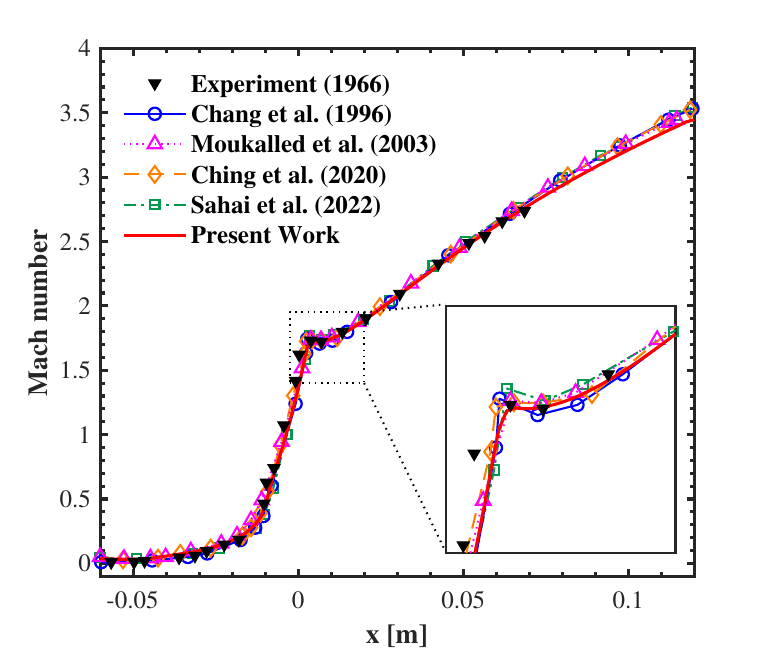}
    \caption{}
    \label{fig:JPL_pure_wall}
\end{subfigure}
\caption{Comparisons of Mach number distributions in the pure gas case. (a) Centerline and (b) nozzle wall.}
\label{Comparison of Mach number in the pure gas}
\end{figure}

Figure \ref{Comparison of Mach number in the pure gas} shows the comparisons of Mach number distributions along the centerline and slip wall for the pure gas case. Along the centerline, the flow accelerates to supersonic speeds as it passes through the nozzle throat ($x=0$), followed by a decrease in Mach number due to the presence of an oblique shock. In contrast, along the wall, the flow continues to accelerate through the expansion region without being affected by the shock. The results demonstrate good agreement with previous numerical and experimental results. 

\begin{figure}[H]
\centering
\begin{subfigure}[H]{0.5\textwidth}
    \centering
    \includegraphics[width=\textwidth]{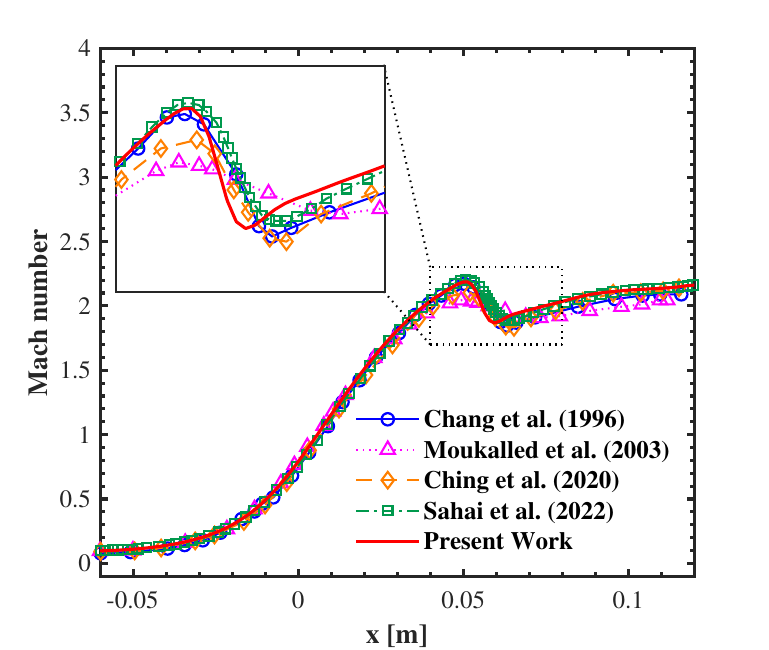}
    \caption{}
    \label{fig:JPL_dust_centerline}
\end{subfigure}%
\hfill
\begin{subfigure}[H]{0.5\textwidth}
    \centering
    \includegraphics[width=\textwidth]{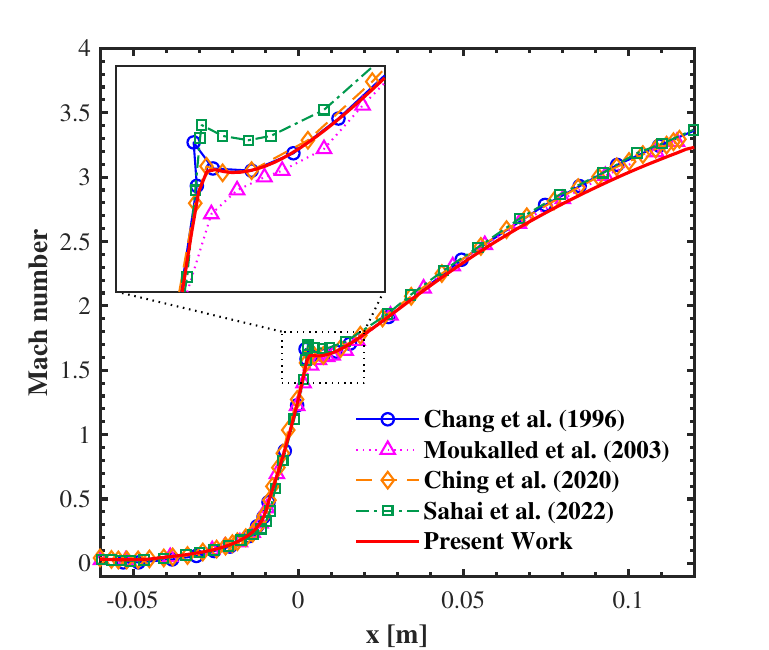}
    \caption{}
    \label{fig:JPL_dust_wall}
\end{subfigure}
\caption{Comparisons of Mach number distributions in the dusty gas case. (a) Centerline and (b) nozzle wall.}
\label{Comparison of Mach number in the dusty gas}
\end{figure}

In the particle-laden simulation, the injection velocity of particles is set equal to the inlet velocity. Wall-particle interaction is modeled as perfectly elastic. Figure \ref{Comparison of Mach number in the dusty gas} presents the results of the dusty gas simulation compared with previous numerical studies \cite{sahai2022variable, ching2020two, chang1996application, moukalled2003pressure}. The particles absorb kinetic energy from the gas phase, leading to flow deceleration and a reduction in shock strength. The particles impact and reflect off the wall in the converging section, pass through the nozzle throat, and remain concentrated near the axis in the diverging section, which is attributed to their slower response to the nozzle flow. As a result, the shock that appears at approximately $x = 0.08~\text{m}$ in the pure gas case is shifted upstream to around $x = 0.05~\text{m}$ in the dusty gas case, with a weakened shock strength. Along the wall, the influence of particles is less significant due to the relatively lower number of particles compared to the centerline. The present result demonstrates good agreement with the predictions from the previous studies.

\begin{figure}[H]
    \centerline{\includegraphics[width=8.0cm]{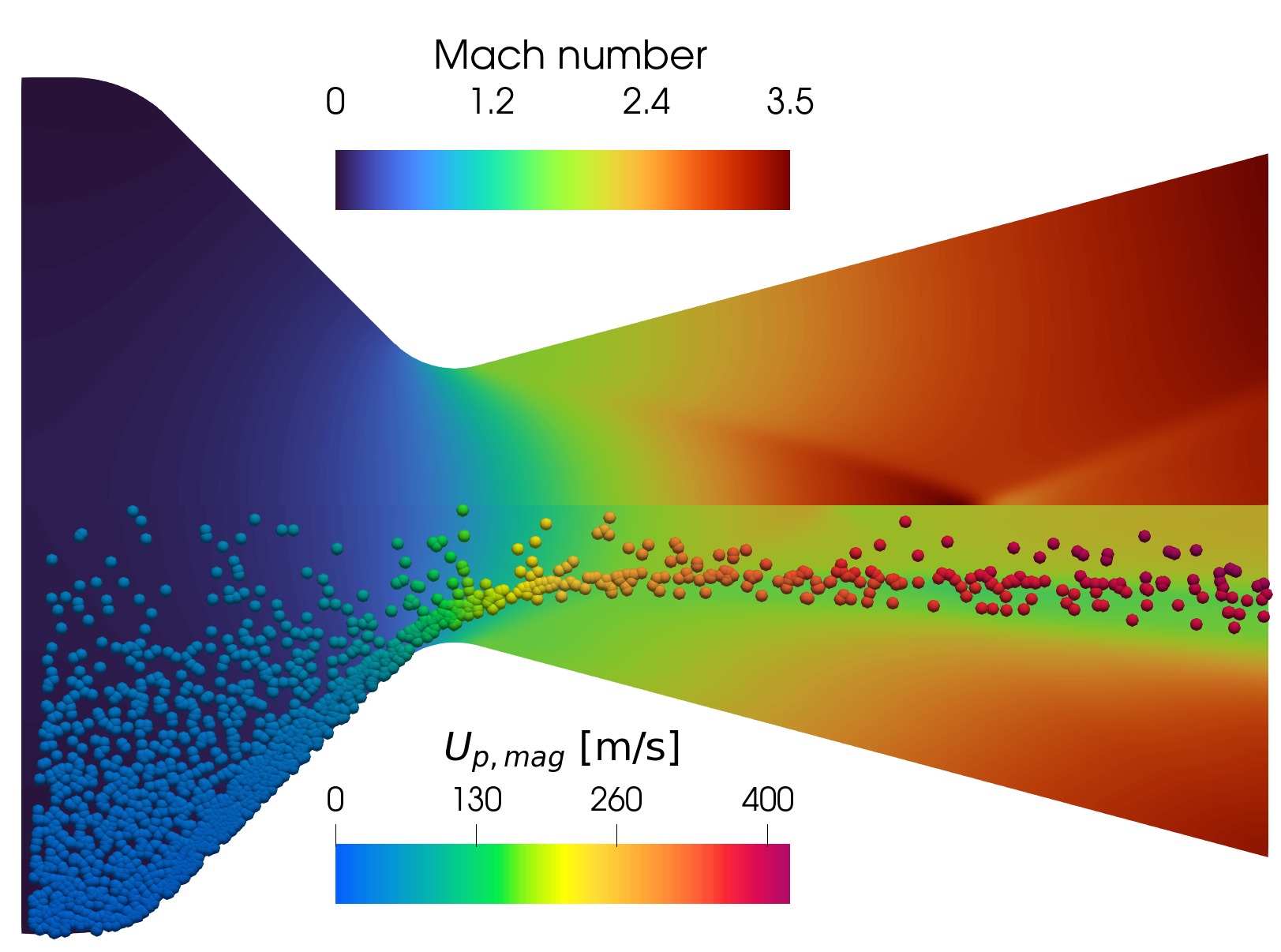}}
    \caption[Comparison of the Mach number of the pure gas(top) and dusty gas(bottom)]{Comparison of the Mach number contours of pure gas (top) and dusty gas (bottom).} \label{Comparison of the Mach number contour}
\end{figure}

Figure~\ref{Comparison of the Mach number contour} compares Mach number contours for the pure gas and dusty gas cases. For the dusty gas case, only a portion of the particles is visualized, with colors indicating the velocity magnitude of the particles. In the expansion region, particles exhibit low sensitivity to the flow, resulting in accumulation near the centerline and a reduced number of particles near the wall. Furthermore, the oblique shock observed in the pure gas appears weakened in the dusty gas.

In summary, the developed Euler–Lagrangian two-way coupling framework accurately captures the gas–particle interactions in the supersonic nozzle flow. It successfully reproduces key features such as shock modification and particle behavior, validating its applicability to dusty gas simulations in high-speed compressible flows.

\vspace{0.5cm}

\subsection{Particle-Induced Surface Effect of Martian Entry Vehicle}
\label{Schiaparelli}
A two-way coupling simulation was conducted for a thermochemical nonequilibrium hypersonic flow around the ExoMars Schiaparelli capsule to further validate the coupling framework. The ExoMars Schiaparelli capsule geometry has a nose radius of 0.6m and a half-cone angle of 70 degrees \cite{palmer2020modeling}. The freestream condition corresponds to the trajectory point at 45 km altitude, and the angle of attack is simplified to 0 degrees to allow for an axisymmetric analysis \cite{hinkle2024efficient}. The Martian atmosphere is modeled using eight species(CO$_2$, N$_2$, CO, NO, O$_2$, C, N, and O), and a two-temperature ($T_{tr}$-$T_{eev}$) model is employed. Energy exchange between two modes is modeled using a Landau-Teller Vibrational-Translational relaxation term with Millikan-White relaxation time \cite{kim2021modification}. The wall boundary condition assumes radiative equilibrium, and the wall temperature is computed accordingly. A fully catalytic wall model for N$_2$ and O$_2$ is applied to account for surface recombination reactions. Particles are modeled based on the dust composition assumed by \citet{papadopoulos1993heatshield}, with a density of 2940 kg/m$^3$ and a latent heat of $8.6 \times 10^6$ J/kg \cite{baldwin1971ablation}. As in the previous study \cite{hinkle2024efficient}, due to the high-temperature conditions in the shock layer, particle vaporization is considered, while the products generated by vaporization are neglected. The drag coefficient and heat transfer models are taken from \citet{henderson1976drag} and \citet{fox1978shock}, respectively. In the particle analysis, two mass loading ratios are considered: (1) 0.0136\% corresponding to the conditions observed during the July 2007 global dust storm by the Mars Reconnaissance Orbiter (MRO) \cite{nmsuMarsAerosol}, and (2) 1.0\% studied by \citet{hinkle2024efficient} to enhance the dust effect by increasing the mass loading by 71 times. The particle size distribution was simplified by using monodisperse particles with a single representative size. Wall-particle interaction is modeled as perfectly non-elastic, and the resulting kinetic energy loss is fully converted to thermal energy and transferred to the wall. A summary of the simulation conditions is presented in Table~\ref{tab:mars-freestream}.

\begin{table}[H]
\caption{Gas and particle phase properties for the ExoMars Schiaparelli case \cite{hinkle2024efficient}.}
\renewcommand{\arraystretch}{0.8}
\centering
\begin{tabular}{cccccc}
\hline\hline
$U_\infty$ (m/s) & $\rho_\infty$ (kg/m$^3$) & $T_\infty$ (K) & $\rho_p$ (kg/m$^3$) & $d_p$ ($\mu$m) \\
\hline
5185 & $2.944 \times 10^{-4}$ & 175 & 2940 & 5.0 \\
\hline\hline
\end{tabular}
\label{tab:mars-freestream}
\end{table}

\begin{figure}[H]
\centering
\begin{subfigure}[t]{0.5\textwidth}
    \centering
    \includegraphics[width=\textwidth]{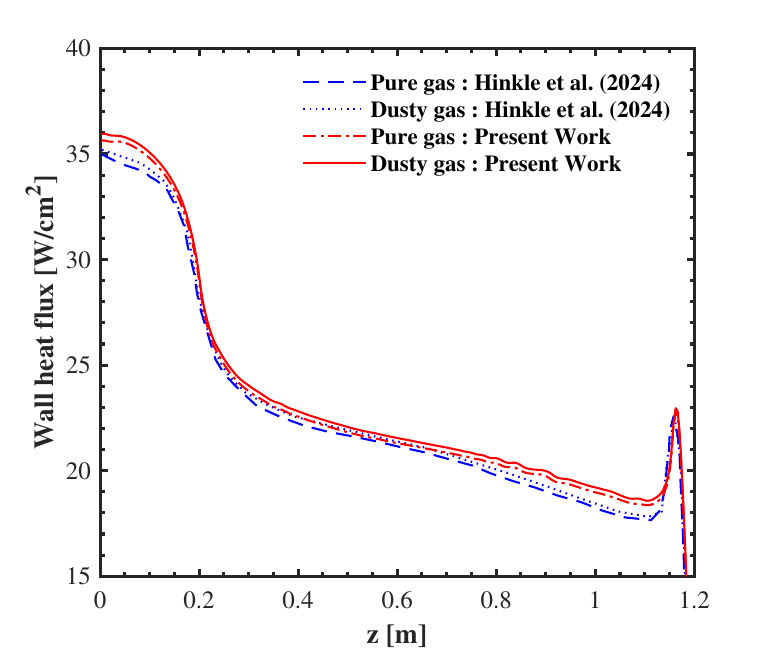}
    \caption{}
    \label{Schiaparelli_heatflux_beta_low}
\end{subfigure}%
\hfill
\begin{subfigure}[t]{0.5\textwidth}
    \centering
    \includegraphics[width=\textwidth]{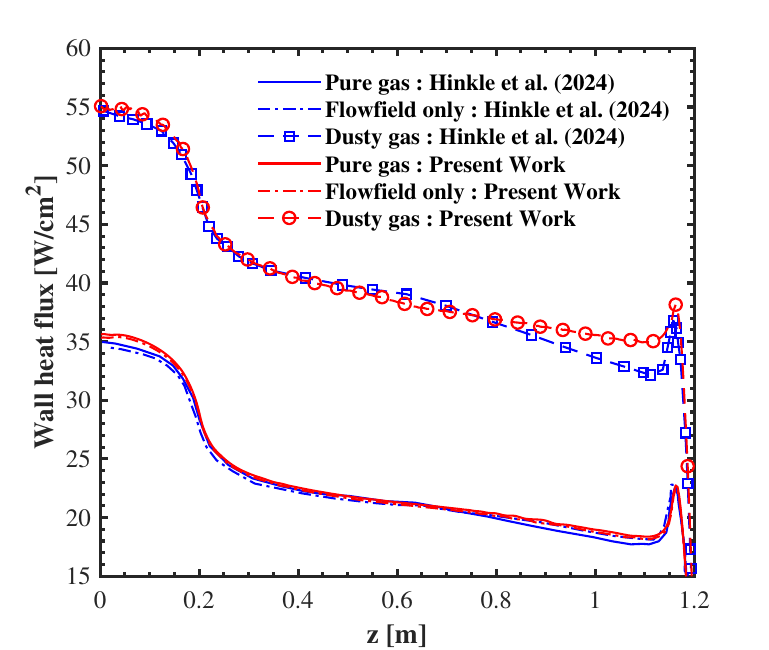}
    \caption{}
    \label{Schiaparelli_heatflux_beta_high}
\end{subfigure}
\caption{Comparisons of the wall heat flux for the ExoMars Schiaparelli case. (a) $\beta$ = 0.014\% and (b) $\beta$ = 1\%.}
\label{fig:Schiaparelli_heatflux}
\end{figure}

Figure~\ref{Schiaparelli_heatflux_beta_low} shows the wall heat flux distribution for the case with a mass loading ratio of 0.0136\%. The "pure gas" case represents the scenario without particles, while the "dusty gas" case includes both convective heat flux and surface collision heating from particles. The results show good agreement with the previous study by Hinkle et al. \cite{hinkle2024efficient}. Under the dust storm condition observed by MRO, the difference in wall heat flux between the pure and dusty gas cases is minimal.

To further evaluate particle effects, Figure~\ref{Schiaparelli_heatflux_beta_high} shows the results for the case with the increased mass loading ratio of 1.0\%. In this case, the "flowfield only" result considers only convective heat flux, while the "dusty gas" result includes both convective heat flux of the gas phase and collision heating. The results show good agreement with the reference data, with minor deviations near the shoulder region. These discrepancies are attributed to differences in the pure gas solution, presumably differences in the thermochemical model. Still, they are limited in magnitude and do not affect the overall recession trend that is the interest of the present study, indicating that the developed framework provides reliable results on the Euler-Lagrangian coupled effects under the thermochemical nonequilibrium two-way coupling scenarios. 

By capturing particle effects in thermochemical nonequilibrium hypersonic flows and reproducing the previous results, the developed two-way coupling framework is validated under these conditions. To quantify the direct thermal influence of particle impacts on the wall, the previously computed collision heat flux was incorporated into the boundary condition that evaluates the wall temperature through the radiative equilibrium wall boundary condition. Specifically, it was applied as an additional source term in the radiative equilibrium wall boundary condition, such that the total wall heat flux was taken as the sum of the convective and particle-induced contributions. This modified boundary condition was then used to evaluate the variation in surface temperature. The results show a clear dependence of the surface temperature on the mass loading ratio. For a mass loading of $0.014\%$ observed by MRO, the surface temperature is nearly identical to that of the pure gas case, whereas a $1\%$ loading causes a significant increase in wall temperature by around 200 K. This demonstrates that the particle mass loading ratio can substantially influence the surface temperature by enhancing the collision-induced heating at the wall.

\begin{figure}[H]
    \centerline{\includegraphics[width=7.0cm]{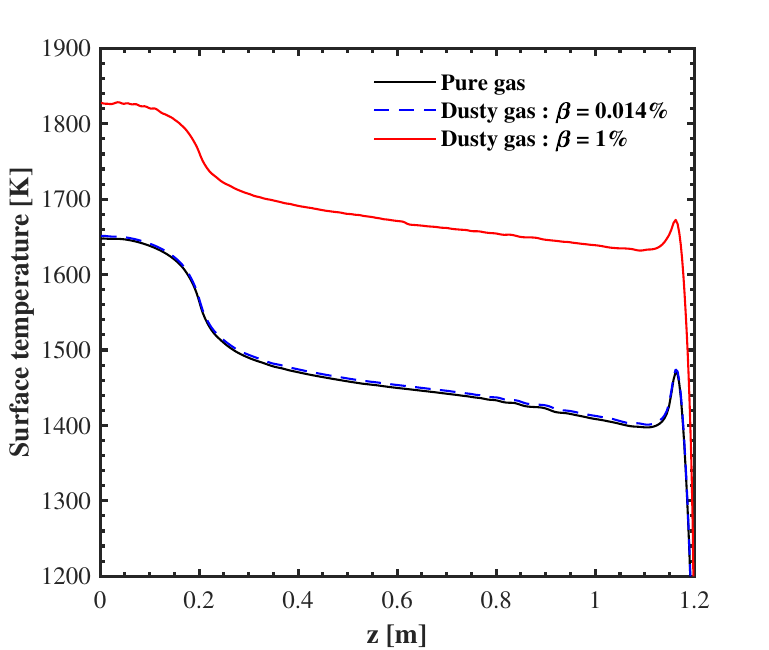}}
    \caption[Effect of the mass loading on the surface temperature]{Effect of the mass loading on the surface temperature.} \label{fig:surface_temp}
\end{figure}
\vspace{-0.5cm}

\begin{figure}[H]    \centerline{\includegraphics[width=7.0cm]{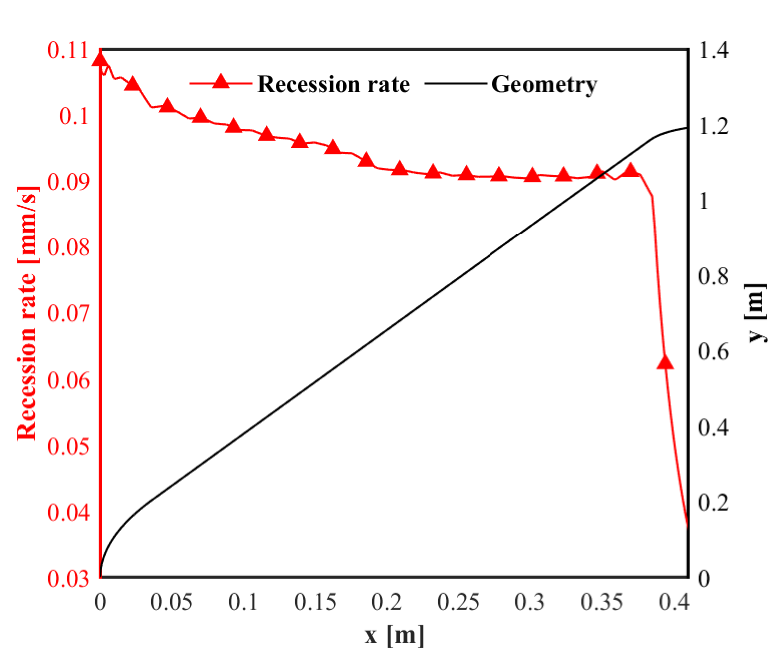}}
    \caption[Surface recession due to particles at $\beta = 0.014\%$]{Surface recession due to particles at $\beta = 0.014\%$.} \label{surface_recess}
\end{figure}

While the thermal impact of dust particles on the wall is relatively small at the MRO measured mass loading ratio, the mechanical damage can be dominant in TPS recession. Therefore, the effect of particle-induced recession is also examined using surface damage modeling. For the MRO observation data-based case with a $0.014\%$ mass loading, the previously described surface damage model is applied to evaluate the recession rate, as shown in Fig.~\ref{surface_recess}. The maximum recession occurs at the stagnation point. The Schiaparelli capsule was coated with Norcoat Liege TPS material, with a thickness of 12.8 mm~\cite{palmer2020modeling}. If exposed to the current entry conditions for approximately 10 seconds, the stagnation point would experience a thickness loss of about 8.4\%. Such a significant thickness loss indicates that, under the realistic assumptions in this case, the particle effects are more critical to the surface recession than to the wall heat flux or the surface temperature.

\section{Reduced-Order Modeling of Cumulative Surface Recession}
\label{Reduced-Order Modeling of Cumulative Surface Recession}
To enable efficient and accurate predictions of particle-induced surface recession during the Schiaparelli atmospheric entry under dust storm conditions, a reduced-order modeling strategy is proposed in this study.
This approach can provide high-fidelity recession data at a significantly reduced computational cost and forms the basis for constructing a surrogate model.
The surrogate model adopts a polynomial chaos expansion (PCE) formulation, which enables the quantification of uncertainty propagation from dust storm parameters to the quantities of interest and facilitates a systematic evaluation of their effects on the aerothermal heating. This section presents the formulation and implementation details of the reduced-order and surrogate modeling framework and proposes analyses of the propagation of parametric model uncertainty.

\subsection{Martian Dust Environment Modeling}

In the Martian atmosphere, particles of various sizes are present, and during dust storms, the particle concentration can increase significantly. In major dust storms, particles can reach altitudes from 50 km to 80 km \cite{majid2016two}, and their distribution varies with altitude. For accurate analysis, the particle size distribution and number density need to be characterized and modeled \cite{ching2021sensitivity}. As discussed by ~\citet{petty2011modified}, the size distribution can be expressed as:

\begin{equation}
    N(r_p) = N_0 \cdot r_p^{\mu} \cdot \exp\left( - \frac{\mu}{\gamma} \left( \frac{r_p}{r_m} \right)^\gamma \right),
\end{equation}

\noindent where $N_0$ is a scaling constant, $r_m$ is the modal radius. $\mu$ and $\gamma$ are shape parameters that define the distribution. The modal radius $r_m$ represents the most probable particle size and controls the relative proportion of small and large particles; a smaller $r_m$ shifts the distribution toward finer particles. $N(r_p)$ denotes the volumetric number density distribution function. Based on observations during the 1971 global dust storm, \citet{toon1977physical} suggested shape parameter values of $\mu = 2$ and $\gamma = 1/2$. Once the modal radius and distribution shape are defined, the number density per unit volume for a given particle radius can be obtained \cite{palmer2020modeling}. To apply this continuous distribution in numerical simulations, it is necessary to discretize the distribution into representative particle sizes, a process known as binning. As noted by  ~\citet{palmer2020modeling}, the particle radius range of \( r_p = 0.5 \, \mu\text{m} \) to \( 9 \, \mu\text{m} \) was divided into 15 discrete bins for computational modeling. Additionally, the mass mixing ratio as a function of altitude was calculated based on data from the July 2007 global dust storm observed by MRO \cite{palmer2020modeling}.

\subsection{Surrogate Modeling of Surface Recession Using Quasi-1D Particle Analysis}

In this section, the cumulative surface recession caused by dust particles along the Martian entry flight trajectory of the ExoMars 2016 Schiaparelli capsule is analyzed. Due to the shortest post-shock distance near the nose, particle-induced recession typically peaks at the stagnation point~\cite{hinkle2022efficient,ching2020two}, resulting in its serving as a quantity of interest in the design phase. Previous studies by \citet{palmer2020modeling} and \citet{hinkle2022efficient} performed three-dimensional simulations along the whole flight trajectory, including angle of attack effects. Both studies predicted the surface recession along the flight trajectory based on the July 2007 global dust storm and reported similar results. Table~\ref{tab:schiaparelli-trajectory} summarizes the freestream conditions and angle of attack at each trajectory point \cite{gulhan2019aerothermal, palmer2020modeling}.

\begin{table}[hbt!]
\caption{Schiaparelli CFD trajectory points \cite{gulhan2019aerothermal,palmer2020modeling}.}
\centering
\begin{tabular}{ccccc}
\hline\hline
Altitude (km) & Density (kg/m$^3$) & Velocity (m/s) & Temperature (K) & Angle of attack (deg) \\
\hline
50.0  & $1.755 \times 10^{-4}$ & 5500.6 & 171.8 & 7.2 \\
45.0  & $2.944 \times 10^{-4}$ & 5185.0 & 175.0 & 7.2 \\
40.0  & $4.825 \times 10^{-4}$ & 4689.0 & 182.4 & 7.2 \\
35.0  & $7.717 \times 10^{-4}$ & 4016.9 & 186.3 & 7.2 \\
30.0  & $1.322 \times 10^{-3}$ & 2913.7 & 190.1 & 6.0 \\
28.2  & $1.542 \times 10^{-3}$ & 2595.4 & 191.6 & 5.8 \\
25.5  & $1.979 \times 10^{-3}$ & 2103.8 & 195.4 & 5.0 \\
23.1  & $2.440 \times 10^{-3}$ & 1570.6 & 199.1 & 4.2 \\
\hline\hline
\end{tabular}
\label{tab:schiaparelli-trajectory}
\end{table}
The flow field along the trajectory was analyzed under thermochemical nonequilibrium assumption using a two-temperature model, consistent with previous works \cite{brandis2022simulation}. However, full 3D simulations for particles require substantial computational cost. Previous studies have shown that particle-induced source terms have a negligible impact on the gas phase in surface recession predictions \cite{ching2020two}, supporting the use of a one-way coupling instead of a two-way framework. Also, the angle of attack has been found to have minimal influence on the surface recession at the stagnation point \cite{ching2020two}. Additionally, particles near the stagnation line travel in nearly straight paths and dominate the nose recession. In axisymmetric flows, particles injected outside the stagnation line are deflected toward the shoulder; their contribution to the stagnation-point erosion is negligible. Therefore, injecting particles only into the first cell along the stagnation line provides a representative estimate of the cumulative effect on the nose recession as the quantity of interest. 

\begin{figure}[H]    \centerline{\includegraphics[width=7.0cm]{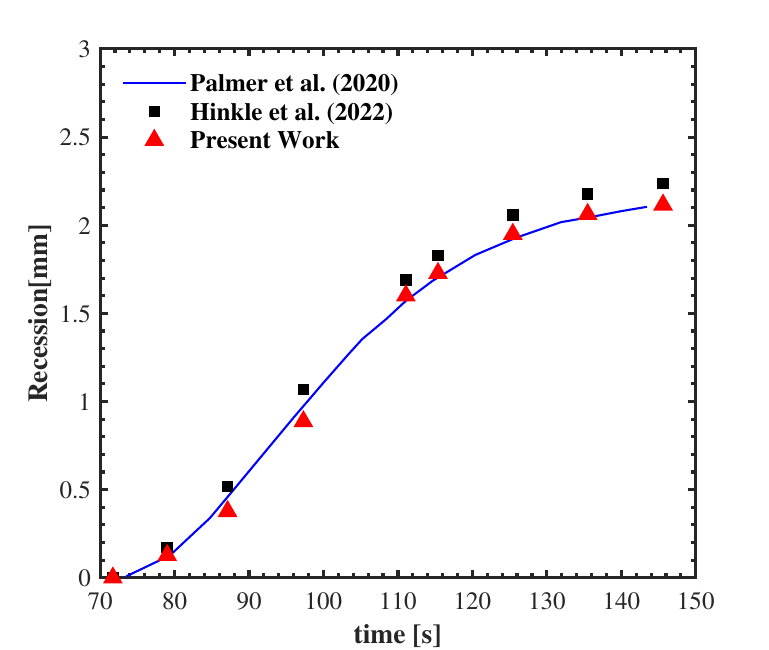}}
    \caption[Surface recession along trajectory at the stagnation-point due to particle impact]{Surface recession along the flight trajectory at the stagnation-point due to particle impact.} \label{fig:traj_recess}
\end{figure}

Based on these insights, a quasi-1D particle analysis is performed by injecting particles only into the first cell along the stagnation line. The surface recession of the Norcoat Liège TPS induced by a single particle is first evaluated using the hemispherical crater model described in Eq.~\eqref{eq:penetration_depth}, and the corresponding particle number flux then scales the result to obtain the cumulative effect. The recession rate $R$ is computed as follows \cite{ching2022simulations}:
\begin{equation}
    R = \sum_l R_l = \sum_l D_{c,l}^2 \mathcal{P}_l \cos{\frac{\pi}{4}} \dot{N}_l,
    \label{eq:recession_model}
\end{equation}

\noindent where $D_{c,l}$ is the crater diameter of particle size group $l$ and $P_l$ represents the penetration depth obtained from Eq.~\eqref{eq:penetration_depth}. The term $\dot{N}_l$ denotes the particle flux per unit area of particle size group $l$. Figure~\ref{fig:traj_recess} presents the simulation results. The predictions of the present study show good agreement with the previous three-dimensional simulations \cite{palmer2020modeling,hinkle2022efficient}, validating the quasi-1D approach. By the end of the trajectory, approximately 18\% of the TPS thickness was found to be lost due to dust-induced erosion.

Based on the proposed quasi-1D framework, a surrogate model for the surface recession rate is constructed using a PCE method. The surrogate model can be expressed as

\begin{equation}
\hat{R}(\boldsymbol{\xi}) = \sum_{i=0}^{P} \alpha_i \Psi_i(\boldsymbol{\xi}),
\label{eq:pce_expansion}
\end{equation}

\noindent where $\hat{R}(\boldsymbol{\xi})$ denotes the surrogate prediction of the surface recession rate corresponding to a given random vector $\boldsymbol{\xi}$, $\Psi_i(\boldsymbol{\xi})$ are the polynomial chaos basis functions, and $\alpha_i$ are the coefficients to be determined. The modal radius and particle density, as identified by \citet{hinkle2022efficient}, are set to the dominant uncertain variables for the stagnation point recession. These two variables were defined with uniform prior distributions, and Legendre polynomials were adopted as their corresponding basis functions. To capture the altitude-dependent variation of surface recession rate, altitude was also included as an additional uncertain input variable. Altitude was modeled using a discrete uniform distribution with discrete Legendre polynomials \cite{neuman1974discrete} as the basis functions. The input parameters and probability distributions are summarized in Table~\ref{tab:pce_inputs}. $\mathcal{U}$ and $\mathcal{U}_{\mathrm d}$ represent the uniform and discrete distribution, respectively. A third-order PCE was constructed, and the surrogate model was trained using 1,000 samples of the quasi-1D model predictions. The polynomial chaos expansion and surrogate construction were implemented using the Chaospy library \cite{feinberg2015chaospy}. To improve predictive accuracy, independent PCE surrogate models were trained separately for low and high altitude regimes across four discrete altitude, as summarized in Table~\ref{tab:pce_inputs}. The \hyperref[Appendix]{Appendix}  provides the basis functions and corresponding expansion coefficients of the trained surrogate, expressed with respect to the input variables.

\begin{table}[t]
\centering
\caption{Input parameter distributions used in the PCE.}
\label{tab:pce_inputs}
\setlength{\tabcolsep}{8pt}
\renewcommand{\arraystretch}{1.2}
\begin{tabular}{lll}
\hline\hline
\textbf{Variable} & \textbf{Symbol [unit]} & \textbf{Distribution} \\
\hline
Modal radius         & $r_m$ [\si{\micro\metre}]                  & $\mathcal{U}(0.3,\,0.5)$ \\
Particle density     & $\rho_p$ [\si{\kilo\gram\per\meter\cubed}] & $\mathcal{U}(2793,\,3087)$ \\
Altitude (low)   & $h$ [\si{\kilo\metre}]                     & $\mathcal{U}_{\mathrm d}(\{23.1,\,25.5,\,28.2,\,30\})$ \\
Altitude (high)  & $h$ [\si{\kilo\metre}]                     & $\mathcal{U}_{\mathrm d}(\{35,\,40,\,45,\,50\})$ \\
\hline\hline
\end{tabular}
\end{table}

\begin{figure}[H]
    \centering
    \begin{subfigure}[b]{0.555\textwidth}
        \centering
        \includegraphics[width=\textwidth]{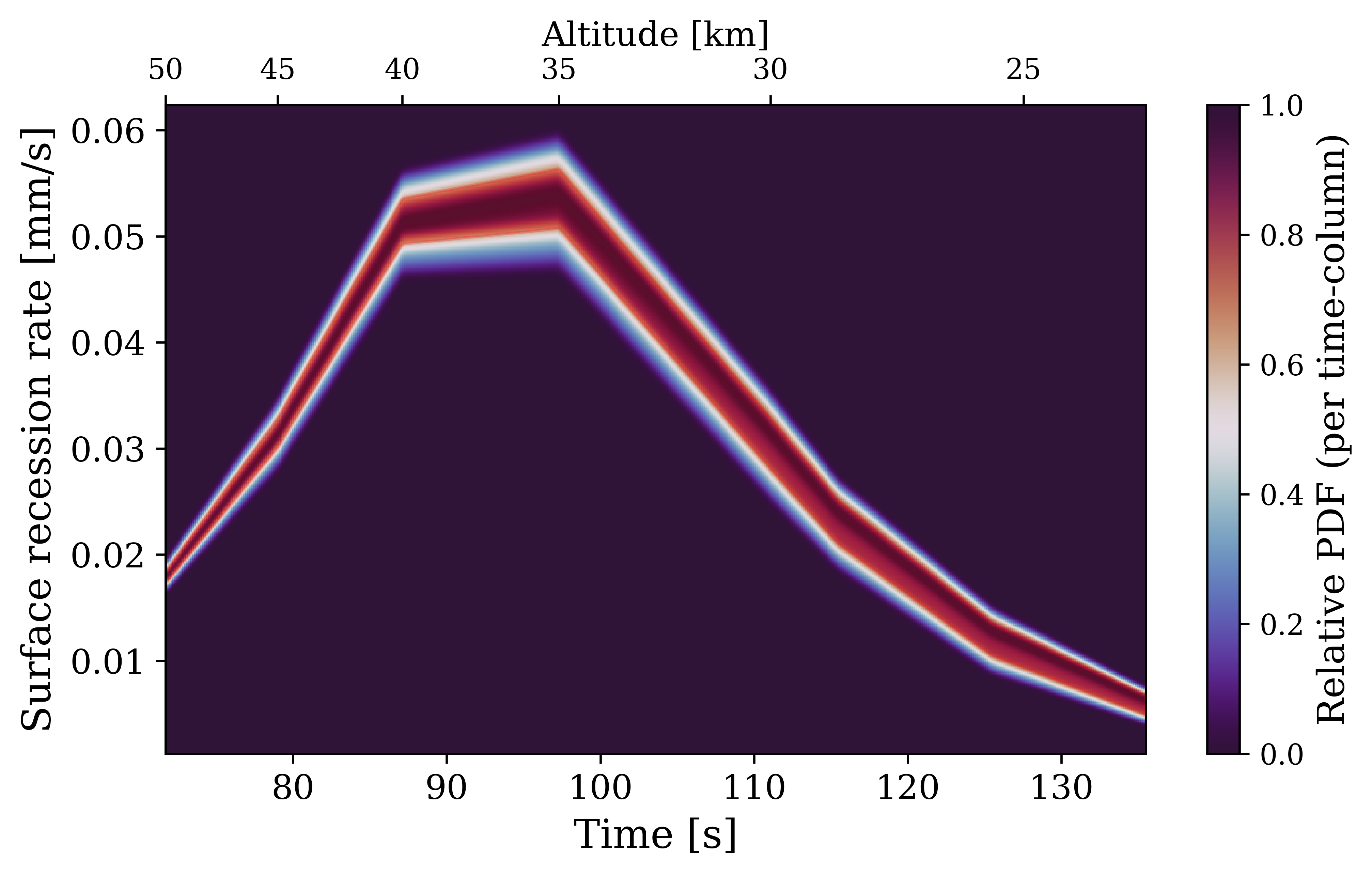}
        \caption[.]{PCE-based probability distribution along the trajectory}
        \label{PCE_rate}
    \end{subfigure}
    \hfill
    \begin{subfigure}[b]{0.43\textwidth}
        \centering
        \includegraphics[width=\textwidth]{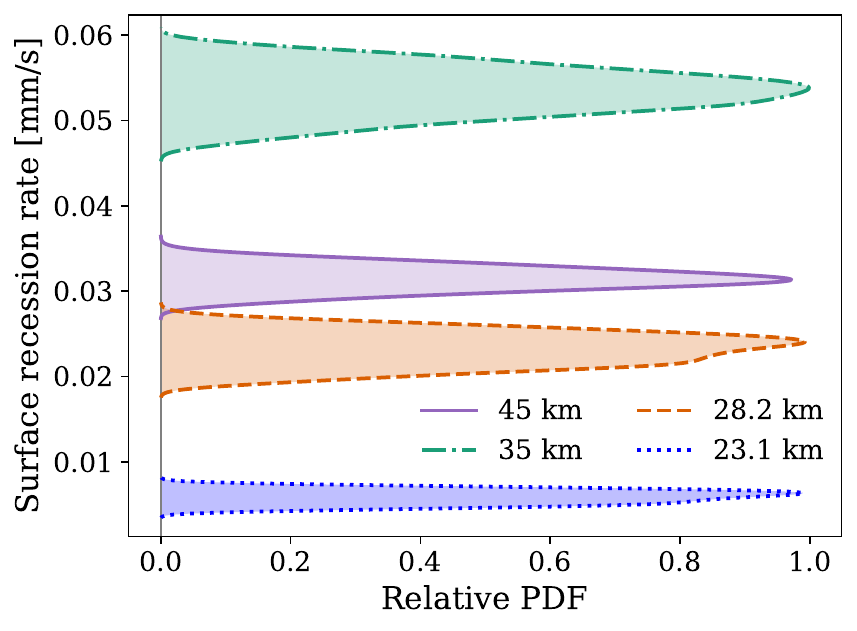}
        \caption[.]{Probability-density slices at the selected altitudes}
        \label{PDF_rate}
    \end{subfigure}
    \caption{PCE results for stagnation-point surface recession rate along trajectory.}
    \label{PCE_and_PDF_rate}
\end{figure}

\begin{figure}[H]
    \centering
    \begin{subfigure}[b]{0.555\textwidth}
        \centering
        \includegraphics[width=\textwidth]{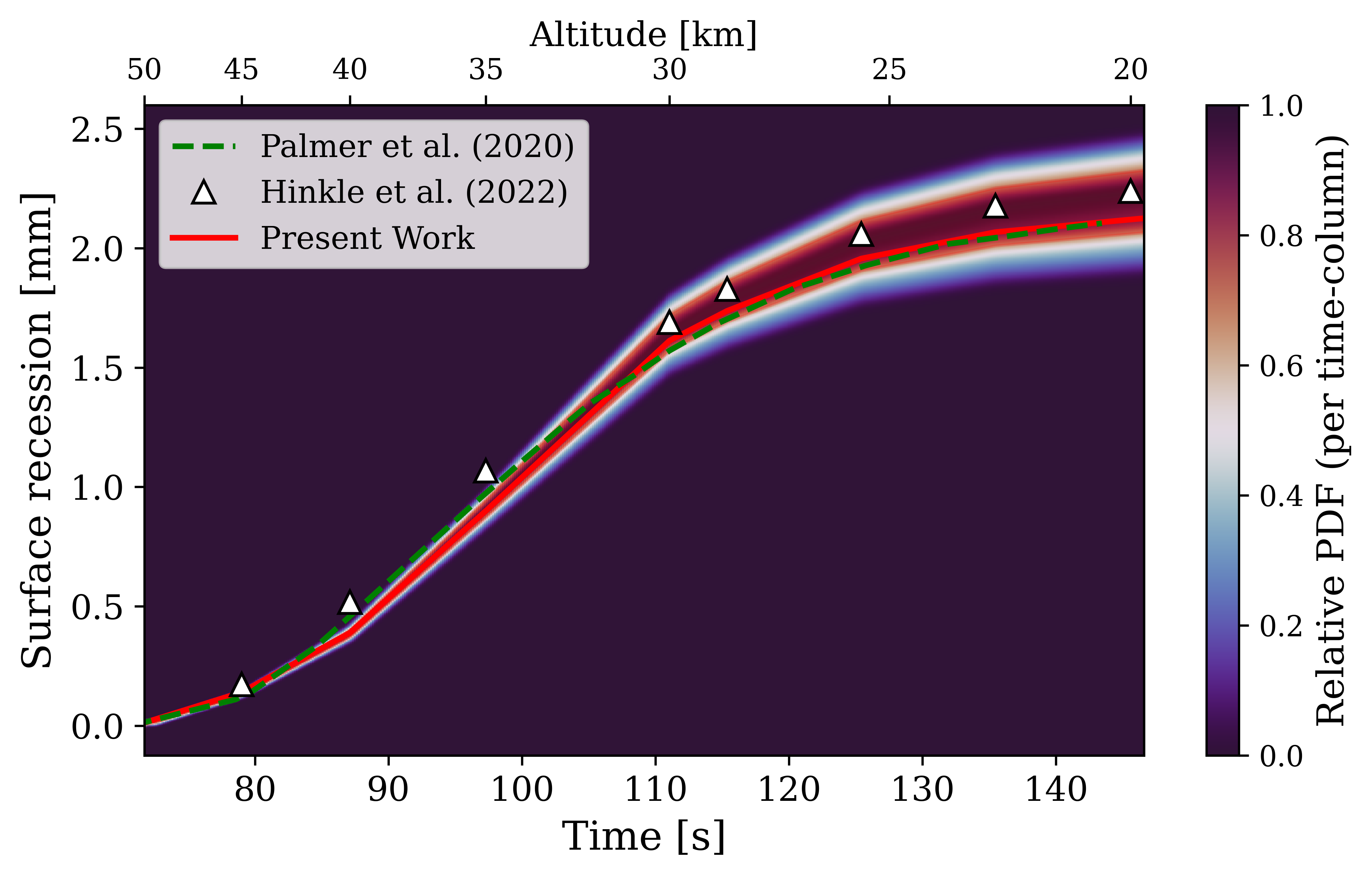}
        \caption[.]{PCE-based probability distribution along the trajectory}
        \label{PCE_depth}
    \end{subfigure}
    \hfill
    \begin{subfigure}[b]{0.43\textwidth}
        \centering
        \includegraphics[width=\textwidth]{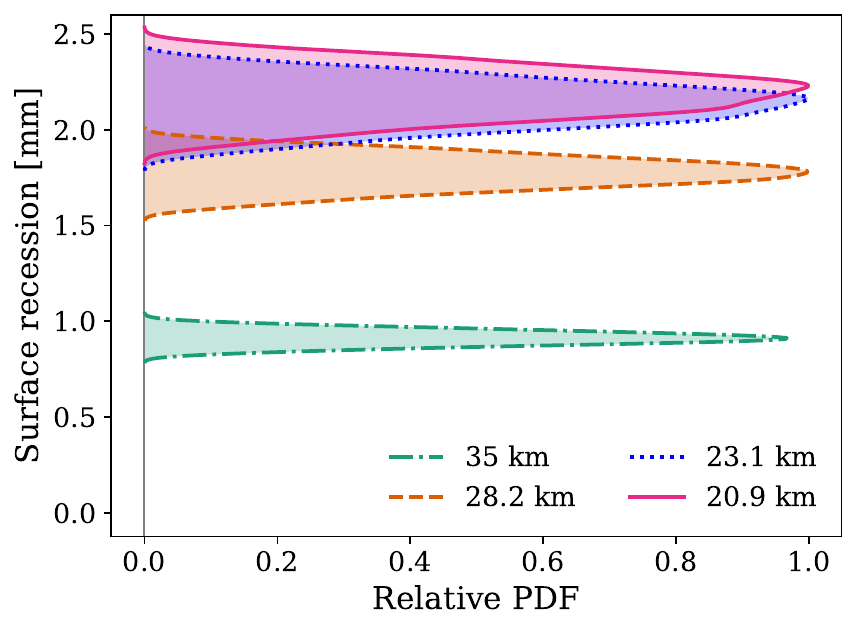}
        \caption[.]{Probability-density slices at the selected altitudes}
        \label{PDF_depth}
    \end{subfigure}
    \caption{Results for stagnation-point surface recession along the trajectory, derived from the PCE predictions.}
    \label{PCE_and_PDF_depth}
\end{figure}

Figure~\ref{PCE_rate} shows the surface recession rate predicted by the PCE surrogate model, while Figure~\ref{PDF_rate} depicts the relative probability density function (PDF) slices at selected altitudes. The PDF shows the likelihood distribution of surface recession based on input parameters. The PDF distribution exhibits the largest variance at 35 km, indicating the highest uncertainty in the predicted surface recession rate at this altitude. The recession rate reaches its maximum in the mid-altitude region (35 km–40 km) and decreases in the low-altitude regions. In the high-altitude region, the particle mass loading ratio is relatively small, resulting in a limited contribution to surface recession. In the low-altitude region, although the particle loading increases, the surrounding gas density becomes sufficiently high that the particle sensitivity to the flow increases, causing deceleration before impact.  As a result, the particle velocity is reduced, leading to a decrease in the recession rate. The cumulative surface recession is obtained by integrating the linearly interpolated recession rate over time. In Figure~\ref{PCE_depth}, the red line (\(r_m = 0.35~\mu\mathrm{m},~\rho_p = 2940~\mathrm{kg/m^3}\)) is compared with previous studies \cite{hinkle2022efficient, palmer2020modeling}, and it shows good agreement with their results. At the same time, the background colormap indicates the relative PDF of recession, indicating the uncertainty resulting from the variations of the particle parameters. Figure~\ref{PDF_depth} presents the relative PDF slices at the selected altitudes. As altitude decreases, particle impacts accumulate, and the amount of recession increases. Because the uncertainty in parameters such as particle size and density accumulates over the trajectory, the standard deviation of the PDF increases with decreasing altitude.
Overall, the proposed quasi-1D framework combined with the PCE surrogate modeling approach successfully reproduces the particle-induced surface recession of the Norcoat Liège TPS under the July 2007 global dust storm conditions. This demonstrates that the proposed approach can be extended to future Martian entry missions under varying dust storm conditions.

\section{Conclusion}
\label{Conclusion}

In this study, an Euler–Lagrangian coupling framework was developed and validated to investigate the effects of solid particles in the high-speed flows. The particle solver ORACLE was coupled with the thermochemical nonequilibrium CFD solver HEGEL and the physico-chemical library PLATO to enable accurate modeling of the particle dynamics, particle–wall interactions, and their couplings with the carrier gas under high-enthalpy conditions. As a result, the proposed coupled framework can simulate a range of high-enthalpy gas conditions, including perfect gas, thermochemical equilibrium, and nonequilibrium conditions, within a single framework. By comparing with the previous studies on the dusty nozzle flow and the ExoMars Schiaparelli atmospheric entry cases, the framework accurately captured particle-induced flowfield changes, surface heating, and surface recession. Under Martian dust storm conditions, the contribution of particles to the wall heat flux and surface temperature increase was shown to be minor under the realistic dust storm environment. In contrast, the surface erosion of the thermal protection system was found to be more significant. 

To reduce the computational cost of full 3D simulations of the dusty hypersonic flows, a quasi-1D analysis method along the stagnation line was proposed. This approach was validated against existing 3D results at a range of angles of attack and provided accurate predictions of the cumulative surface recession at the substantially reduced computational cost. The recession at the stagnation point was estimated to reduce the TPS thickness by approximately 18\% by the end of the entry trajectory. Using this quasi-1D framework, a PCE surrogate model was also proposed to propagate the uncertainties in key particle parameters. The PCE model accurately reproduced the recession rate and resultant time-dependent surface recession, and quantified the probabilistic distribution of recession. The parametric uncertainties in surface recession increased at lower altitudes due to increased particle impacts.

Overall, the developed framework offers an efficient and reliable tool for analyzing particle-laden hypersonic flows and their effects on the aerothermal environments. The approach using the PCE surrogate model for particle-induced surface recession based on the quasi-1D analysis can quantify the uncertainty impacts and offer insights for TPS design under dusty atmospheric conditions in future planetary entry missions.

\section*{Acknowledgements}
This work was partially supported by the AFOSR under Award No. FA9550-25-1-0119 (to S.M.J.). The authors would like to thank Dr. A. Sahai (NASA Ames Research Center), Dr. A. Hinkle (NASA Langley Research Center), and Dr. S. Hosder (Missouri University of Science and Technology) for useful discussions on Lagrangian approaches. The authors also thank Dr. A. Munafò and Dr. M. Panesi (University of California, Irvine) for providing access to the PLATO and HEGEL software.

\section*{Appendix}
\label{Appendix}
This section describes the details of the PCE surrogate model with input parameters \((r_m,\rho_p,k)\). The continuous inputs $r_m$ and $\rho_p$ are defined over the interval $[a,b]$ in which $a$ and $b$ respectively stand for the minimum and maximum values of the given parameter and are assumed to follow a uniform distribution. To construct the polynomial basis on this interval, the Legendre polynomials are used as follows:

\begin{equation}
\psi_{0}(\xi) = 1, \quad
\psi_{1}(\xi) = \xi-m, \quad
\psi_{2}(\xi) = (\xi-m)^{2} - \tfrac{h^2}{3}, \quad
\psi_{3}(\xi) = (\xi-m)^{3} - \tfrac{3}{5}h^2(\xi-m),
\label{eq:monic_legendre_general}
\end{equation}
\noindent where $m = (a+b)/2$ and $h = (b-a)/2$. The symbol $\xi$ denotes the input variable (\emph{i.e.}, $r_m$ and $\rho_p$), and the subscript $n$ in $\psi_n$ indicates the polynomial order.
\newpage
To account for the variation of surface recession with altitude, it is incorporated as a discrete variable. Four representative altitudes were selected and mapped to an integer index \(k \in \{1,2,3,4\}\), ordered from highest to lowest altitude (\(k=1\) corresponds to the highest altitude). The discrete Legendre polynomials for this index provide the basis functions \cite{neuman1974discrete}:

\begin{equation}
\chi_{0}(k)=1,\qquad
\chi_{1}(k)=k-\tfrac{5}{2},\qquad
\chi_{2}(k)=k^{2}-5k+5,\qquad
\chi_{3}(k)=k^{3}-\tfrac{15}{2}k^{2}+\tfrac{167}{10}k-\tfrac{21}{2}.
\label{eq:discretelegendre}
\end{equation}

\noindent
The expansion coefficients \(\alpha_{a,b,c}\) associated with the basis 
\(\Psi_{a,b,c}=\psi_a(r_m)\psi_b(\rho_p)\chi_c(k)\) are listed in 
Table~\ref{tab:coeff}, which summarizes the third-order PCE surrogate. This table includes the coefficients obtained from both the high-altitude and low-altitude PCE, allowing the prediction of the particle recession rate cumulative over the flight time.

\begin{table}[t]
\centering
\caption{PCE coefficients for $(r_m,\rho_p, k)$.}
\begin{tabular}{c c c c c}
\hline
Term & $(a,b,c)$ & Basis term $\Psi_{a,b,c}$ & Low alt. & High alt. \\
\hline
1 & (0,0,0) & $\psi_{0}(\xi_m)\psi_{0}(\xi_\rho)\chi_{0}(k)$ & 1.792e-02 & 3.860e-02 \\
2 & (0,0,1) & $\psi_{0}\psi_{0}\chi_{1}$ & -8.528e-03 & 1.271e-02 \\
3 & (0,1,0) & $\psi_{0}\psi_{1}\chi_{0}$ & 7.785e-06 & 1.254e-05 \\
4 & (1,0,0) & $\psi_{1}\psi_{0}\chi_{0}$ & 2.486e-02 & 2.231e-02 \\ 
5 & (0,0,2) & $\psi_{0}\psi_{0}\chi_{2}$ & 2.659e-04 & -2.867e-03 \\
6 & (0,1,1) & $\psi_{0}\psi_{1}\chi_{1}$ & -3.053e-06 & 4.492e-06 \\
7 & (0,2,0) & $\psi_{0}\psi_{2}\chi_{0}$ & -1.852e-10 & -3.579e-10 \\
8 & (1,0,1) & $\psi_{1}\psi_{0}\chi_{1}$ & -7.223e-03 & 1.014e-02 \\
9 & (1,1,0) & $\psi_{1}\psi_{1}\chi_{0}$ & 4.526e-06 & 4.920e-06 \\
10 & (2,0,0) & $\psi_{2}\psi_{0}\chi_{0}$ & -4.852e-02 & -4.303e-02 \\
11 & (0,0,3) & $\psi_{0}\psi_{0}\chi_{3}$ & 1.445e-03 & -4.022e-03 \\
12 & (0,1,2) & $\psi_{0}\psi_{1}\chi_{2}$ & -1.480e-07 & -7.063e-07 \\
13 & (0,2,1) & $\psi_{0}\psi_{2}\chi_{1}$ & -5.817e-11 & 5.238e-11 \\
14 & (0,3,0) & $\psi_{0}\psi_{3}\chi_{0}$ & 4.196e-13 & -2.961e-12 \\
15 & (1,0,2) & $\psi_{1}\psi_{0}\chi_{2}$ & -1.227e-03 & 2.745e-05 \\
16 & (1,1,1) & $\psi_{1}\psi_{1}\chi_{1}$ & -1.320e-06 & 1.774e-06 \\
17 & (1,2,0) & $\psi_{1}\psi_{2}\chi_{0}$ & -2.934e-09 & -4.008e-09 \\
18 & (2,0,1) & $\psi_{2}\psi_{0}\chi_{1}$ & 1.570e-02 & -2.096e-02 \\
19 & (2,1,0) & $\psi_{2}\psi_{1}\chi_{0}$ & -1.126e-05 & -2.689e-06 \\
20 & (3,0,0) & $\psi_{3}\psi_{0}\chi_{0}$ & 7.510e-02 & 6.382e-02 \\
\hline
\label{tab:coeff}
\end{tabular}
\end{table}

\bibliography{mybib}

\end{document}